\documentclass[amsmath,amssymb]{revtex4}

\usepackage{graphicx}
\usepackage{dcolumn}
\usepackage{bm}
\usepackage{mathrsfs}
\usepackage{appendix}
\usepackage{bbm}
\usepackage{color}

\def\be{\begin{equation}}
\def\ee{\end{equation}}
\def\bd{\begin{displaymath}}
\def\ed{\end{displaymath}}
\def\-{\phantom{-}}

\begin{document}
%
%

\title{Spin Rotation Technique for Non-Collinear Magnetic Systems: Application to the Generalized Villain Model}

\author{J. T. Haraldsen and R. S. Fishman}

\affiliation{Materials Science and Technology Division, Oak Ridge National Laboratory, Oak Ridge, Tennessee 37831, USA}

\date{\today}

\begin{abstract}

This work develops a new generalized technique for determining the static and dynamic properties of
any non-collinear magnetic system. By rotating the spin operators into the local spin reference frame, 
we evaluate the zeroth, first, and second order terms in a Holstein-Primakoff expansion, and through a Green's
functions approach, we determine the structure factor intensities for the spin-wave
frequencies. To demonstrate this technique, we examine the spin-wave dynamics of the
generalized Villain model with a varying interchain interaction. The new interchain coupling expands the
overall phase diagram with the realization of two non-equivalent canted spin configurations.
The rotational Holstein-Primakoff expansion provides
both analytical and numerical results for the spin dynamics and intensities of these phases.

\end{abstract}

\maketitle

\section{Introduction}

\begin{figure}
\includegraphics[width=5.0in]{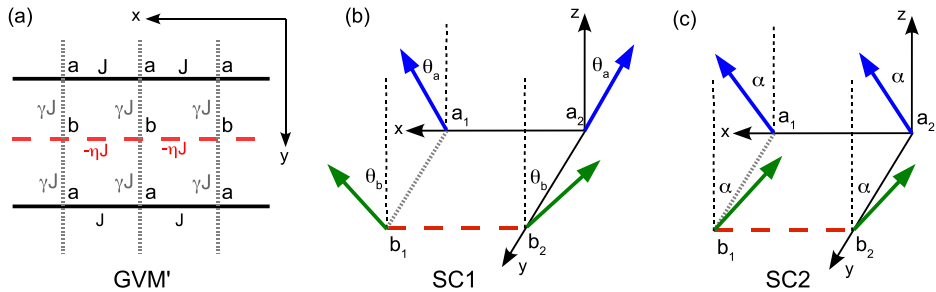}
\caption{(a) The generalized Villain model (GVM$^{\prime}$) with Heisenberg interactions $J$,
$\gamma J$, and $-\eta J$. (b) The GVM$^{\prime}$ with anti-parallel spin configuration (SC1), where
the local moments of sites
$a$ and $b$ reside in the $xz$-plane and are described by the angles ($\theta_a$,$\theta_b$). (c) The GVM$^{\prime}$ with parallel
spin configuration (SC2),
where the local moments of sites $a$ and $b$ reside in the $xz$-plane and are canted by the angle $\alpha$. Note that SC2 with $\alpha$ = $\pi$/2 is rotationally equivalent to
SC1 with (0, $\pi$).}
\label{VillainModel}
\end{figure}

Disordered and frustrated magnetic systems have provided condensed-matter physics
with a range of
complicated problems \cite{kit:87}. Magnetic
frustration is typically induced by competing nearest and next-nearest neighbor 
antiferromagnetic (AF) interactions \cite{kit:87,vil:77}. To minimize the energy, 
the local moments frequently rotate into a non-collinear ground-state configuration \cite{vil:77,ber:86}. 
Theoretical and
experimental efforts have worked to identify and characterize non-collinear
magnetic systems. Quantum Monte Carlo and Hartree-Fock
calculations have been used to describe canted antiferromagnetic (CAF) phases in double
quantum dots and bi-layer
quantum Hall systems \cite{khr:00,san:01}, while $ab~initio$ methods have been used to
examine non-collinearity in magnetic atomic chains \cite{laz:04}. Through neutron-diffraction
studies, layered
borocarbide systems $R$B$_2$C
($R$ = Dy, Ho, and Er) have demonstrated conventional and
unconventional magnetic
correlations \cite{dui:03}. Other systems like the cuprate, ruthenate, and
manganite systems (where the
competition between AF and ferromagnetic (FM) order has been examined in
great depth \cite{dag:01,ada:00,koo:01}) also display non-collinear
characteristics \cite{cof:91,she:93,jor:01,cao:01,nak:02,dag:01,ada:00,koo:01}.  To help understand
non-collinear magnetic systems,
we present a new technique for modeling the static and dynamic
properties of
canted local moments in any periodic system.

After rotating the spin operators into the local reference frame at each site, we use 
a Holstein-Primakoff (HP) expansion to determine the classical spin energy and
the spin-wave (SW) frequencies. By solving the equations-of-motion for coupled 
Green's functions, we determine the structure factor (SF) intensities for any 
eigenfrequency of the system. While a rotation into the local reference frame of the 
moments has been employed before to describe the SW modes of spin glasses and 
other non-collinear two-dimensional AFs \cite{wal:80,sas:83,sas:92,car:04}, the general technique introduced 
in this paper can 
easily describe any magnetic system, with arbitrary exchange coupling and
single-ion anisotropy.  The inputs into this method are the interaction parameters and
moment angles;  the outputs are the SW frequencies and intensities. 

To demonstrate this general technique, we investigate the generalized Villain model
with an added interchain coupling. In recent years, the generalized Villain model (GVM) has
been frequently used to test approaches to study magnetically frustrated systems and to
help understand more complicated magnetic
systems similar to those mentioned above \cite{sas:92,fis:04a,fis:04b,fis:04c}. In the GVM, chains of FM interactions $J$ and AF interactions $-\eta J$
along the $x$-axis are coupled together in the $y$ direction by
$J$ \cite{sas:92,fis:04a,fis:04b,fis:04c}, as illustrated in Fig. \ref{VillainModel}(a). The original Villain model assumed 
that
$\eta  = 1$, which denotes full frustration \cite{vil:77}.
This model was generalized to $\eta~\ne$ 1 by Berge $et~al.$ \cite{ber:86}, accounting for coupling of different magnitudes within the FM and AF chains. Within the GVM, a CAF phase is stabilized for $\eta > \eta_c$. The critical value increases with applied magnetic field from a value of $\eta_c$=$\frac{1}{3}$ in zero field. In 1992, Saslow and Erwin \cite{sas:92} employed a HP expansion to numerically examine the SW and SF
intensities for this generalized model. 

In this paper, the GVM is extended
further by introducing a variable $y$-axis coupling $\gamma J$, as
shown in Fig. \ref{VillainModel}(a). We call this model the 
GVM$^{\prime}$.  Therefore, the GVM$^{\prime}$ with $\gamma$ = 1
reduces to the GVM discussed previously \cite{sas:92,fis:04a,fis:04b,fis:04c}. As in the GVM, all interactions are confined to the
$xy$ plane but the magnetic is applied along the $z$ axis.
Through an examination of the classical energies, we demonstrate that there exists two separate canted
spin configurations throughout \{$\eta$,$\gamma$\} phase space. Figures \ref{VillainModel}(b) and (c)
show the possible canted spin configurations within the GVM$^{\prime}$. In Fig. \ref{VillainModel}(b), 
the local moments are canted in the $xz$-plane by angles $\theta_{a}$ and $\theta_{b}$. In this configuration, the spins projected onto the $xy$ plane are antiparallel on sites $a_1$ and $a_2$ as well as on sites $b_1$ and $b_2$ \cite{sas:92,fis:04a,fis:04b,fis:04c}. This canted spin configuration occurs in the GVM with $\gamma$ = 1. With the introduction of the variable $\gamma$,  a new spin configuration arises:
the corresponding projected
moments are parallel with moment angles ($\alpha$), as shown in Fig. \ref{VillainModel}(c). 
For notational
convenience, we call these SC1 (Spin Configuration 1) for the anti-parallel case (Fig. \ref{VillainModel}(b)) and SC2
(Spin Configuration 2) for the parallel case (Fig. \ref{VillainModel}(c)).

Through an examination of the classical limit, we determine the phase diagram within the parameter space \{$\eta$,$\gamma$\}. The GVM$^{\prime}$ is found to support three phases (FM, SC1, and SC2). Using the rotational Holstein-Primakoff expansion, the SW frequencies are determined analytically
 for all phases. The SF intensities for the FM and zero-field SC2 phases are determined exactly; the rest are solved numerically.

\section{The General Rotation Model}

As described in Ref. \cite{fis:04b}, the Hamiltonian for canted magnetic systems
can be simplified by rotating into the reference frame for each moment:
$\mathbf{\bar{S}}_i$ = $\underline{U}_i \mathbf{S}_i$, where $\underline{U}_i$ is the unitary rotation matrix
for site $i$ (discussed in Appendix A). In the classical limit, $\mathbf{\bar{S}}_i$ points along its local $z$-axis. 
The general Hamiltonian
is given by
\be
\begin{array}{ll}
H &\displaystyle = -\frac{1}{2}\sum_{i \neq j} J_{ij} \mathbf{{S}}_i \cdot \mathbf{{S}}_j -
\sum_i D_i\mathbf{{S}}_{iz}^2 - B \sum_i \mathbf{{S}}_{iz} \\ \\
& \displaystyle = -\frac{1}{2}\sum_{i \neq j} J_{ij} \mathbf{\bar{S}}_i \cdot \underline{U}_i \underline{U}_j^{-1}\mathbf{\bar{S}}_j -
\sum_i D_i(\underline{U}_i^{-1}\mathbf{\bar{S}}_i)_z^2 - B \sum_i (\underline{U}_i^{-1}\mathbf{\bar{S}}_i)_z, \\
\label{genH}
\end{array}
\ee
where $\mathbf{S}_i$ are the local moments for site $i$ , $J_{ij}$ is the interaction between
sites $i$ and $j$, $D_i$ is the single-ion anisotropy, and $B$ is the applied magnetic field. It should be noted that $\underline{U}_i$ only depends on the moment angles of the spins at site $i$.
 The canting of a local moment can be 
described as a rotation by $\theta$ in the $xz$-plane, with another rotation by $\psi$ in the $xy$-plane. Therefore,
each local moment can be described by the Euler angles $\theta$ and $\psi$ \cite{arf:01}.
The Hamiltonian is expanded in powers of 1/$\sqrt{S}$ about the classical or high-spin limit:
$H$=$E_{0}$+$H_{1}$+$H_{2}$+.... Within the HP formalism, the spin operators in the local reference frames
become: $\bar{S}_{iz} = S -a_{i}^{\dag} a_{i}$, $\bar{S}_{i+} = \sqrt{2S}a_{i}$ and
$\bar{S}_{i-} = \sqrt{2S}a_{i}^{\dag}$ \cite{kit:87}. The zeroth-order $E_0$ term corresponds to the classical energy and the second-order term $H_2$ describes the dynamics of non-interacting SWs. The first-order 
term $H_1$ vanishes when the local moments minimize the classical energy $E_0$ for a specific 
interaction pair $\eta$ and $\gamma$. Each term, up to second order, is discussed futher below. Higher 
order terms correspond to SW interactions that are unimportant at low temperatures
and for small 1/S. 

\subsection{Zeroth Order: Classical Energy}

From the above Hamiltonian, the zeroth-order
terms describe the classical energy and can be written as
\be
E_{0} = -\frac{1}{2}\sum_{i,j} J_{ij} S_iS_j F_{zz}^{ij} - \sum_i D_iS_i^2 \mathrm{cos}(\theta_i)^2 -
B^{\prime} \sum_i S_i^2 \mathrm{cos}(\theta_i),
\label{zerothH}
\ee
where $S_i$ is the moment magnitude for site $i$, $B^{\prime}$ $\equiv$ $B/JS$, and $F_{zz}^{ij}$ is a
rotation coefficient given by the angle rotation matrix in Appendix A. It only depends on the angles of the
spins on sites $i$ and $j$. From an examination of
the global minima, the appropriate spin configuration
can be determined by minimizing this classical energy.

\subsection{First Order: Linear Terms}

The HP expansion produces terms that are linear with respect to the creation and
annihilation operators. Therefore, the 
first-order Hamiltonian is given by
\be
H_{1} = -\sum_{i,j} \frac{S_iS_j}{\sqrt{2}}J_{ij}  \Big( \frac{1}{\sqrt {S_i}}(F_1^{ij} a_{i}^{\dag} + F_1^{ij*} a_{i}) +\frac{1}{\sqrt{S_j}}(F_2^{ij} a_{j}^{\dag} + F_2^{ij*} a_{j})\Big)
+ \sum_{i}  \sqrt{\frac{S_i^3}{2}}\big(B^{\prime}+2D_i\sin(\theta_i)\cos(\theta_i)\big)\Big(a_{i}^{\dag} + a_{i}\Big),
\ee
where $F_1^{ij}$ = $F_{xz}^{ij}$ + $iF_{yz}^{ij}$ and $F_2^{ij}$ = $F_{zx}^{ij}$ + $iF_{zy}^{ij}$.
Here, $F_{xz}^{ij}$, $F_{zx}^{ij}$, $F_{yz}^{ij}$, and $F_{zy}^{ij}$ are rotation 
coefficients that depend only on the
moment angles for the interacting spins on sites $i$ and $j$ (described in
Appendix A). The
linear terms correspond to the creation and annihilation of SW's from the vacuum.
Assuming the system is in a proper ground state
with the angles that minimize the energy, the nonphysical first-order coefficients of $a_{i}$ and
$a_{i}^{\dag}$ must vanish for each spin site.

\subsection{Second Order: Spin Dynamics and Structure Factor Intensities}

The second-order terms in the HP expansion describe the spin
dynamics. We Fourier transform the spin operators by:
$a_{\mathbf{k}}^{(r)} = 1/N\sum_i^{(r)} e^{-i\mathbf{k}\cdot\mathbf{R_i}}a_i$
and
$a_{\mathbf{k}}^{(r)\dag}= 1/N\sum_i^{(r)} e^{i\mathbf{k}\cdot\mathbf{R_i}} a_i^{\dag} $,
where the sums are restricted to sub-lattice (SL) $r$. For each SL of the system, the moment angles are the same. 
The second-order
terms of the generalized Hamiltonian can then be written as
\be
\begin{array}{c}
H_{2} = \displaystyle \sum_{r,s}\sum_{u,\mathbf{k}} z_{rs}^{(u)}J_{rs}^{(u)} \sqrt{S_rS_s}
\Big\{ 
2G_1^{rs}\Gamma_{\mathbf{k}}^{rs(u)}a_{\mathbf{k}}^{(r)\dag}a_{\mathbf{k}}^{(s)}
+ G_2^{rs} \Gamma_{\mathbf{k}}^{rs(u)} a_{\mathbf{k}}^{(r)} a_{\mathbf{-k}}^{(s)} +
G_2^{rs*} \Gamma_{\mathbf{k}}^{rs(u)*} a_{\mathbf{k}}^{(r)\dag} a_{\mathbf{-k}}^{(s)\dag} +
 F_{zz}^{rs} \big(a_{\mathbf{k}}^{(r)\dag} a_{\mathbf{k}}^{(r)} \\ \\ \displaystyle + a_{\mathbf{k}}^{(s)\dag} a_{\mathbf{k}}^{(s)} \big)\Big\}
-\frac{1}{2} \sum_{r,\mathbf{k}} D_rS_r\Big(\sin(\theta_r)^2 \big(a_{\mathbf{k}}^{(r)\dag} a_{\mathbf{-k}}^{(r)\dag} +
a_{\mathbf{k}}^{(r)} a_{\mathbf{-k}}^{(r)} + a_{\mathbf{k}}^{(r)} a_{\mathbf{k}}^{(r)\dag} +
a_{\mathbf{k}}^{(r)\dag} a_{\mathbf{k}}^{(r)} \big)  - 4\cos(\theta_r)^{2} a_{\mathbf{k}}^{(r)\dag} a_{\mathbf{k}}^{(r)}\Big)\\ \\
\displaystyle - B^{\prime}\sum_{r,\mathbf{k}} S_r \mathrm{cos}(\theta_r) a_{\mathbf{k}}^{(r)\dag} a_{\mathbf{k}}^{(r)},
\end{array}
\label{GenHam}
\ee
where $\mathbf{k}$ is the momentum vector. Here, $z_{rs}^{(u)}$ is the number of SL $s$ sites coupled by the interaction $J_{rs}^{(u)}$ to a site on SL $r$, and $u$ denotes the multiple possible interactions from SL $r$ to SL $s$. For example, a FM with a single SL can have both nearest-neighbor and next nearest-neighbor interactions with $u$ = 1 and 2. We have also defined $\Gamma_{\mathbf{k}}^{rs(u)}$ = $1/z_{rs}^{(u)}\sum_{\mathbf{d}}e^{-i\mathbf{k}\cdot\mathbf{d}^{(u)}}$ with $\mathbf{{d}}^{(u)} = \mathbf{R}_j - \mathbf{R}_i$ where $\mathbf{R}_i$ on SL $r$ and $\mathbf{R}_j$ on SL $s$ are coupled by the exchange $J_{rs}^{(u)}$. Note that $\Gamma_{\mathbf{k} = 0}^{rs(u)}$ = 1 and  $\Gamma_{\mathbf{k}}^{rs(u)}$ = $\Gamma_{-\mathbf{k}}^{rs(u)*}$ = $\Gamma_{\mathbf{k}}^{sr(u)*}$. Finally, $G_1^{rs}$ and $G_2^{rs}$ are rotation coefficients that depend only on the
moment angles for the specific SL (described in
Appendix A). 

To determine the SW frequencies $\omega_{\mathbf{k}}$, we solve the equation-of-motion for the vectors $\mathbf{v_{k}} = [a_{\mathbf{k}}^{(1)},
a_{\mathbf{k}}^{(1)\dag},a_{\mathbf{k}}^{(2)},
a_{\mathbf{k}}^{(2)\dag},...,a_{\mathbf{k}}^{(s_t)},
a_{\mathbf{k}}^{(s_t)\dag}]$, which may be written in terms of the 2$s_t$ x 2$s_t$ matrix $ \underline{M}(\mathbf{k})$ as 
\be
id\mathbf{v_k}/dt = -\big[ \underline{H}_{2},\mathbf{v_k}\big] =  \underline{M}(\mathbf{k}) \mathbf{v_k},
\ee
where $s_t$ is the number of spin SLs. The SW frequencies are
then determined from the condition Det[$  \underline{M}(\mathbf{k}) - \omega_{\mathbf{k}} \underline{I}$] = 0, where only positive frequencies are retained.

The structure factor in a magnetic system describes the intensity expected from 
experiment for SW modes with resolution-limited width \cite{squ:78}. In the case of a standard FM with identical nearest-neighbor interactions, the structure
factor is constant throughout $k$-space. However, as the spins cant,
the wave-vector dependence becomes important. 

Local stability in a magnetic system requires two conditions: 1) All SW frequencies must be real for every $\mathbf{k}$ and 2) the SW weights $W_{\mathbf{k}}^{(t)}$ must be positive. The weights are given by the coefficients of the delta functions in the spin-spin correlation function
\be
S(\mathbf{k},\omega) = \frac{1}{2}\Big[S^{+,-}(\mathbf{k},\omega)+S^{-,+}(\mathbf{k},\omega)\Big]+S^{z,z}(\mathbf{k},\omega) = \sum_{t} W_{\mathbf{k}}^{(t)}\delta(\omega-\omega_{\mathbf{k}}^{(t)}),
\ee
where $\omega$ is the eigenfrequency for the spin-wave \cite{squ:78,bal:89} and the sum over $t$ means a sum over all SW modes. The total number of transverse or longitudinal SW modes in the first Brillouin zone equals the number of magnetic SLs $s_t$.  

Generally, the spin-spin correlation function is
\be
S^{\alpha,\beta}(\mathbf{k},\omega) = \frac{1}{ N} \int dt~ e^{-i \omega t}
\sum_{i,j} e^{i \mathbf{k\cdot (\mathbf{R}_j-\mathbf{R}_i)}} \big< \mathbf{S}_i^{\alpha} (0)\mathbf{S}_j^{\beta} (t)\big> ,
\ee
where $\alpha$, $\beta$ = $+$, $-$, and $z$ \cite{squ:78,bal:89} The transverse terms ($\alpha$,$\beta$ = $+$, $-$) correspond to $S^{+,-}(\mathbf{k},\omega)$ and $S^{-,+}(\mathbf{k},\omega)$ in $S(\mathbf{k},\omega)$, while 
the longitudinal term $S^{z,z}(\mathbf{k},\omega)$ with $\alpha$ = $\beta$ = $z$ is only nonzero 
when the system is canted. 
Here, $\big< \mathbf{S}_i^{\alpha} (0)\mathbf{S}_j^{\beta} (t)\big>$ = 
$\big< \left(\underline{U}_i^{-1}\mathbf{\bar{S}}_i (0)\right)^{\alpha}\left(\underline{U}_j^{-1}\mathbf{\bar{S}}_j (t)\right)^{\beta}\big>$ 
is rotated from the local-moment frame to the global frame.
By expanding and solving for the spin Green's functions, we can write $S^{\alpha,\beta}(\mathbf{k},\omega)$ as
\be
S^{\alpha,\beta}(\mathbf{k},\omega) = -\frac{4}{\pi} \lim_{\delta\rightarrow 0} \rm{Im} \left(Tr\left(\underline{G}(\mathbf{k},\omega + i\delta) \underline{C}_{\alpha\beta} \right) \right) ,
\ee
where 
\be
\underline{G}(\mathbf{k},\omega + i\delta) = \int_0^{1/T} d\tau~e^{i\omega_l \tau} \underline{g}^\mathbf{k}(\tau) \Bigr\vert_{i\omega_l\rightarrow\omega + i\delta},
\ee
$\omega_l = 2l\pi T$,
and $C_{\alpha \beta}^{nm}$ are the rotational coefficients for the canted moments, which depend only on the angles $\theta$ and $\psi$ (described in Appendix A).  Here $n$ = 2$r$ -1, 2$r$ for SL $r$ and
\be
g_{nm}^\mathbf{k}(\tau) = -\big< T_{\tau} \mathbf{v}_{n\mathbf{k}}(\tau) \mathbf{v}_{m\mathbf{k}}\big>,
\ee
where $v_{n\mathbf{k}}(\tau) = e^{H_2\tau}v_{n\mathbf{k}}e^{-H_2\tau}$.
Using the equations-of-motion of $\mathbf{v}_{n\mathbf{k}}(\tau)$, the Green's function matrix can be solved as
\be
G_{nm}(\mathbf{k},\omega + i\delta) =\left[\frac{-1}{(\omega + i\delta) \underline{I} -  \underline{M}(\mathbf{k})} \underline{N}\right]_{nm},
\ee
where $ \underline{I}$ is an 2$s_t$ x 2$s_t$ identity matrix and $ \underline{N}$ is an 2$s_t$ x 2$s_t$ matrix defining the commutation
relations such that $[\mathbf{v}_{n,\mathbf{k}},\mathbf{v}_{m,\mathbf{k}}] = N_{nm}$.  

From the commutation relation, $[S_i^+,S_j^-] = 2 S_{z} \delta_{ij}$, one
can determine the net magnetic moment from the sum rule 
\be
\big<S_z\big> = \int_{-\infty}^{\infty}d\omega S_z(\mathbf{k},\omega),
\label{sumrule}
\ee 
where $S_{z}(\mathbf{k},\omega)$ = $\frac{1}{2}\big(S^{+-}(\mathbf{k},\omega)$-$S^{-+}(\mathbf{k},\omega)\big)$. Note that $S_{z}(\mathbf{k},\omega)$ is not the same as the longitudinal spin-spin correlation function, $S^{z,z}(\mathbf{k},\omega)$. Since $\big<S_z\big>$ does not depend on $\mathbf{k}$, neither does the right-hand side of Eq. (\ref{sumrule}).

\section{The Generalized Villain Model: Classical Energy and Phase Boundaries}

To demonstrate the rotational technique described above, we examine the
GVM$^{\prime}$ described in Fig. \ref{VillainModel}.
This model neglects anisotropy, but includes an applied magnetic field.
The introduction of $\gamma$ into the GVM$^{\prime}$ expands the overall interaction phase 
space, which can illuminate some of the interesting phenomena
seen within the cuprate and manganite systems. An overall analysis of the numerical
and analytical results will provide detailed information about the GVM$^{\prime}$, as well as a unique look at the
nature of the spin phases as one moves through the interaction phase space. 

While the general rotational
Hamiltonian describes interactions between spin sites
with moment rotations in both the $xz$ ($\theta$) and $xy$ ($\psi$) planes, the
GVM$^{\prime}$ constrains the moment angles to rotations in the $xz$-plane, which
greatly simplifies the technique. Therefore, spin-spin interactions can be denoted as angle pairs ($\theta_a$,$\theta_b$). This constraint also allows the SW frequencies for the GVM$^{\prime}$ to be determined
analytically.

\begin{figure}
\begin{center}
\includegraphics[width=4.5in]{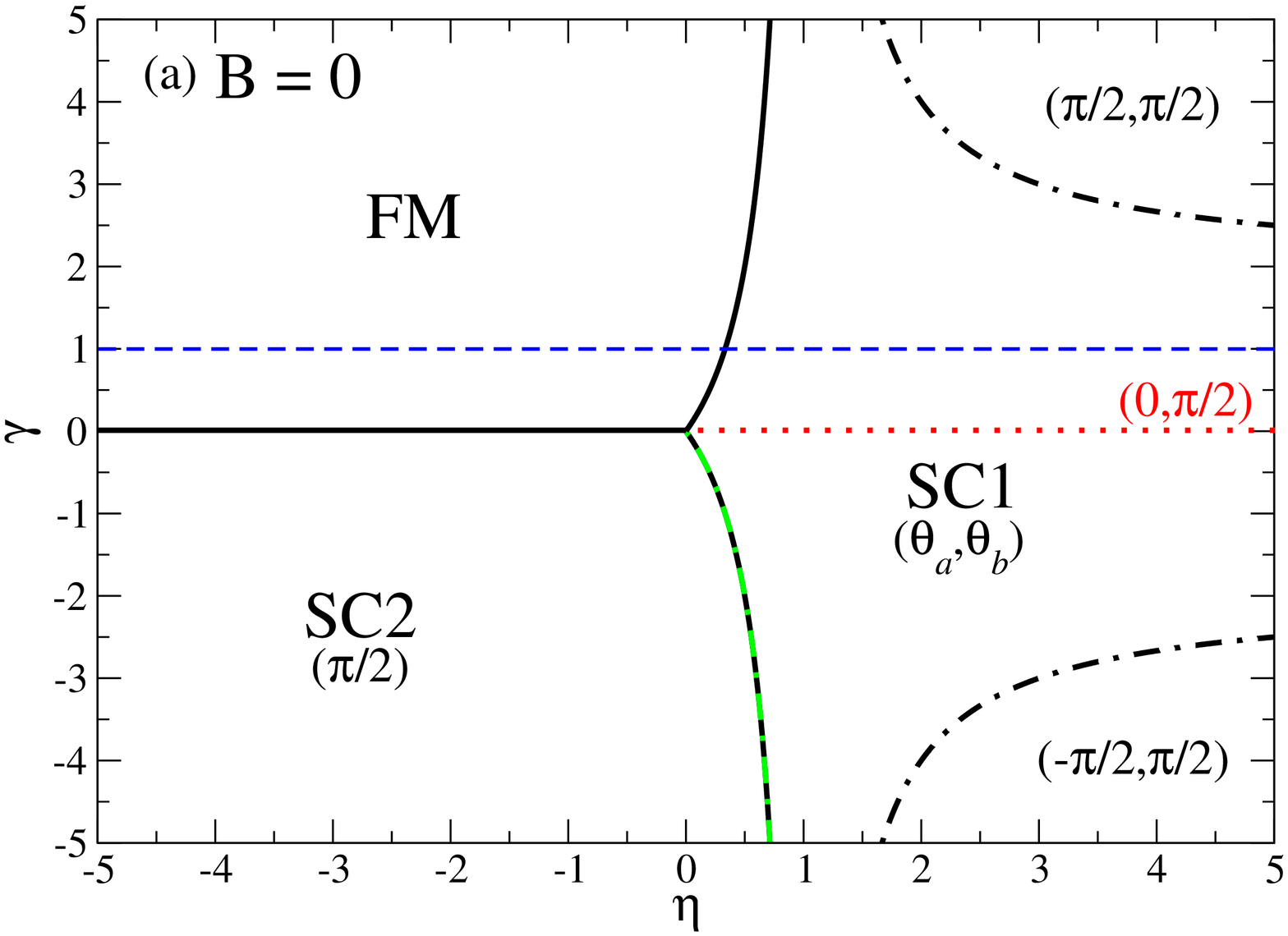}
\includegraphics[width=4.5in]{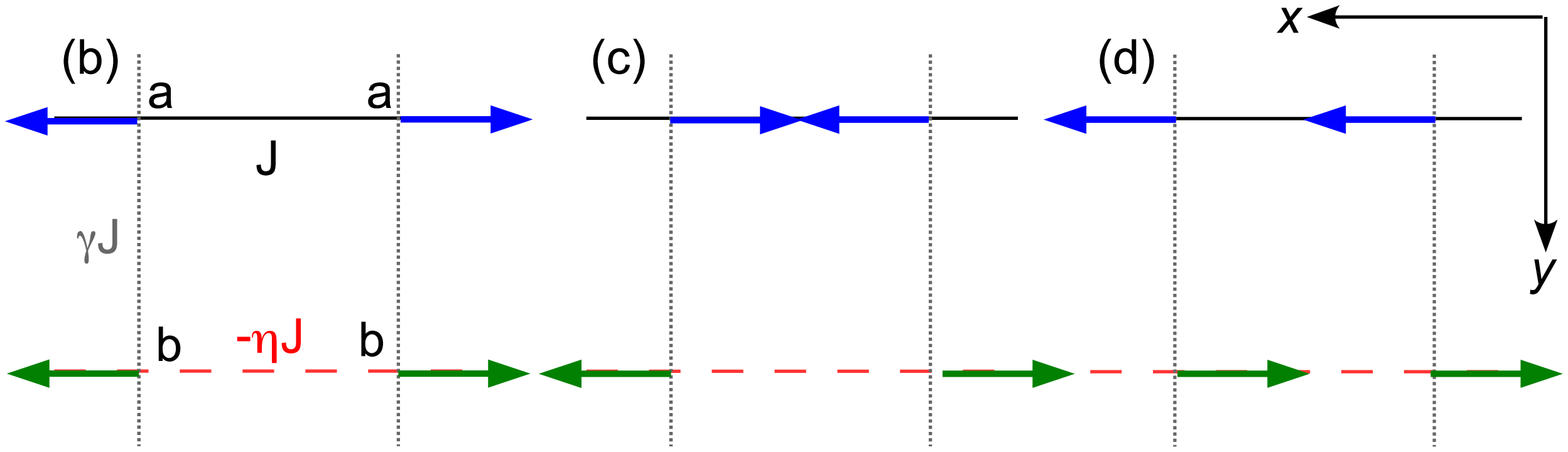}
\caption{(a) The phase diagram for the GVM$^{\prime}$ at $B$ = 0 in the interaction space of
$\gamma$ and $\eta$. The green/black dotted line indicates the phase boundary
between SC1 and SC2 and the blue dashed
line shows the sub-space of the GVM$^{\prime}$ ($\gamma$ = 1). (b)-(d) Pictorial
representations of the (b) ($\pi$/2,$\pi$/2) and (c) (-$\pi$/2,$\pi$/2) limit regions (black dash-dotted line) of the SC1 phase, and (d) the SC2 phase at zero field.}
\label{VillainPhase}
\end{center}
\end{figure}

Through an examination of the classical energy, we determine the energy 
boundaries for each spin configuration. From Eq. (\ref{zerothH}), the classical energy of SC1
is
\be
E_{0}^{(\rm{SC1})} = \frac{JS^2}{2}\Big(\eta \mathrm{cos}(2\theta_b)-\mathrm{cos}(2\theta_a) -
2\gamma \mathrm{cos}(\theta_a - \theta_b)
- B^{\prime}(\mathrm{cos}(\theta_a) + \mathrm{cos}(\theta_b)\Big),
\label{zerothperp}
\ee
Minimizing $E_{0}^{(\rm{SC1})}$ with respect to $\theta_a$ and $\theta_b$ yields
the relations
\be
\begin{array}{c}
\displaystyle \mathrm{sin}(2\theta_a) + \gamma \mathrm{sin}(\theta_a - \theta_b) +
\frac{1}{2}B^{\prime}\mathrm{sin}(\theta_a)=0,\\ \\
\displaystyle \eta \mathrm{sin}(2\theta_b) + \gamma \mathrm{sin}(\theta_a - \theta_b) -
\frac{1}{2}B^{\prime}\mathrm{sin}(\theta_b)=0.\\
\end{array}
\label{Eminperp}
\ee
Assuming that ($\theta_a$,$\theta_b$) satisfy these criteria, the first-order
terms from the Hamiltonian $H_{1}^{(\rm{SC1})}$ will vanish. At zero field and in the limit of large $\eta$, the 
local moments cant toward the angles ($\tan^{-1}(\gamma/\sqrt{4-\gamma^2})$,$\pi$/2).
 In the limit $\gamma\rightarrow$1,
Eqs. (\ref{zerothperp}) and (\ref{Eminperp}) agree with the results of Ref. [\onlinecite{fis:04b}],
where $\theta_a$ and $\theta_b$ approach angles smaller than $\pi/6$ and $\pi/2$, respectively. 
With increasing $\eta$ and $|\gamma|$, the zero-field SC1 phase approaches one of
the planar phases shown in Fig. \ref{VillainPhase}(b) and (c).

By linearizing Eq. (\ref{Eminperp}), we obtain the phase boundary between the SC1 and FM
phases :
\be
B^{\prime} = 2\Big(\eta - \gamma - 1  \pm \sqrt{\gamma^2 +(\eta+1)^2}\Big).
\label{critfield}
\ee
In the limit $\gamma\rightarrow$1, this gives the relation obtained by Gabay $et~al.$ \cite{gab:89}. 
Solving Eq. (\ref{critfield})
for $\gamma$, we find the phase boundaries
\be
\gamma = \pm \frac{(4+B^{\prime})(4\eta-B^{\prime})}{4(2-2\eta+B^{\prime})}.
\label{FM-CAF}
\ee
Figure \ref{VillainPhase}(a) shows the phase diagram for the zero-field GVM$^{\prime}$. Here, the
blue dotted line goes along $\gamma$ = 1, where the critical value
$\eta_c$ = 1/3, is consistent with previous work \cite{ber:86}.

\begin{figure}
\includegraphics[width=4.5in]{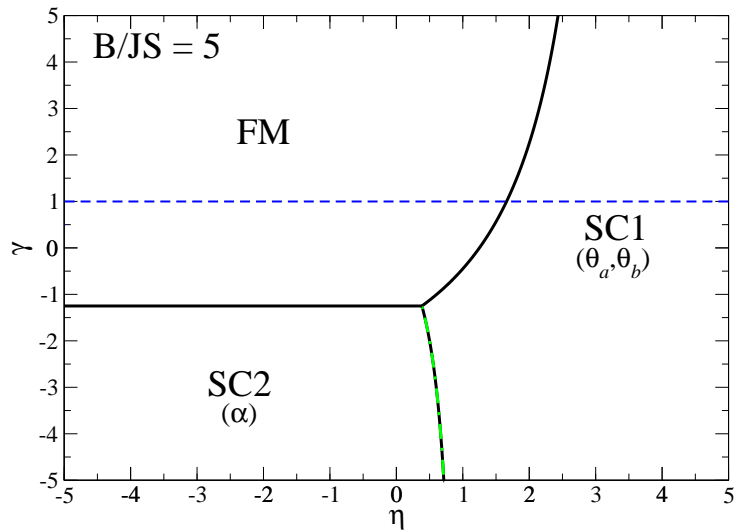}
\caption{Phase Diagram for the GVM$^{\prime}$ at $B^{\prime}$ = $B/JS$ = 5 in the interaction space of $\gamma$
and $\eta$. The green/black dotted line indicates the phase boundary between
SC1 and SC2. The blue dashed line shows the sub-space
of the GVM$^{\prime}$ where $\gamma$ = 1.} 
\label{VillainPhaseB5}
\end{figure}

In zero field, SC1 with angles (0,$\pi$) is equivalent to SC2 with $\alpha$ = $\pi/2$. However, due to the absences of
anisotropy in the GVM$^{\prime}$, a magnetic field in the $z$ direction immediately produces a spin flop 
into the $xy$ plane with the spins canted towards the $z$-axis. This configuration corresponds to SC2 with $\alpha~<$ $\pi/2$.  Figure \ref{VillainPhase}(d) shows the SC2 phase at zero field, where $\alpha$ = $\pi$/2 
throughout the region.
Equation (\ref{FM-CAF}) gives the field dependence of the SC1/FM boundary.

To find the field-dependent boundaries for the SC2 phase, we examine the classical energy
\be
E_{0}^{(\rm{SC2})} = \frac{JS^2}{2}\Big( \eta-1 - 2\gamma \mathrm{cos}(2\alpha)
- 2 B^{\prime}\mathrm{cos}(\alpha)\Big).
\label{zerothpara}
\ee
At zero field, SC2 has the same energy as
SC1. However, SC2 has the
lower energy when a magnetic field is applied, so it does not follow
the same field dependence as SC1. By minimizing the energy, we obtain the SC2 angle
\be
\alpha = \tan^{-1}\left[\frac{\sqrt{16\gamma^2-B^{\prime}}}{4\gamma}\right],
\label{alpha}
\ee
which depends
only on $B^{\prime}$ and $\gamma$. Consequently, the SC2/FM phase boundary
is given by
$B^{\prime}= -4\gamma$ and the SC1/SC2 phase boundary is given by
\be
\gamma = -\frac{2\eta}{(\eta-1)},
\label{AF-CAF}
\ee
which is shown by the green/black dashed lines in Figs. \ref{VillainPhase}(a)
and \ref{VillainPhaseB5}. This reveals that the SC1/SC2 boundary does not depend on field.
As the magnetic field increases, the FM phase expands along this boundary, creating a
triple point at $\eta = B^{\prime}/(B^{\prime}+8)$ and $\gamma=-B^{\prime}/4$.
The expansion of the FM phase space with field is clearly
illustrated in Fig. \ref{VillainPhaseB5}, where $B^{\prime}$ = 5.

\section{Generalized Villain Model: Spin-Waves and Structure Factors}

Using the methods described in section 2, we apply this new technique to the spin dynamics of the
GVM$^{\prime}$ in the three regions of the phase diagram 
(FM, SC1, and SC2). Since the environments surrounding the moments at sites $a_1$ and $a_2$ are equivalent, they can be considered part of the same sublattice.  Similarly, for sites $b_1$ and $b_2$.
The SW frequencies can be determined
analytically for all phases. The SF intensities will be obtained
for a more complete picture of the system. In the FM and zero-field SC2 phases, we
solve the SF intensities analytically. However, due to the added complexity of the nonzero-field SC2
and SC1 phases, those intensities must be determined numerically.

The second order GVM$^{\prime}$ Hamiltonian can be written as
\be
H_{2} = JS \sum_{\mathbf{k},r,s} \Big( a_{\mathbf{k}}^{(r)\dag}
a_{\mathbf{k}}^{(s)}A_{\mathbf{k}}^{(r,s)} + (a_{\mathbf{-k}}^{(r)}a_{\mathbf{k}}^{(s)}+
a_{\mathbf{-k}}^{(r)\dag}a_{\mathbf{k}}^{(s)\dag}) B_{\mathbf{k}}^{(r,s)}\Big),
\label{GVMHam}
\ee
where $A_{k}^{(r,s)}$ and $B_{k}^{(r,s)}$ are coefficients that
describe the interactions while $r$, $s$ = $a$ or $b$ represent the two SLs \cite{fis:04b}.
The resulting SW frequencies
can be expressed analytically using the modified 
coefficients for the different spin configurations, with expressions given in Appendix B.

\begin{figure}
\includegraphics[width=6.0in]{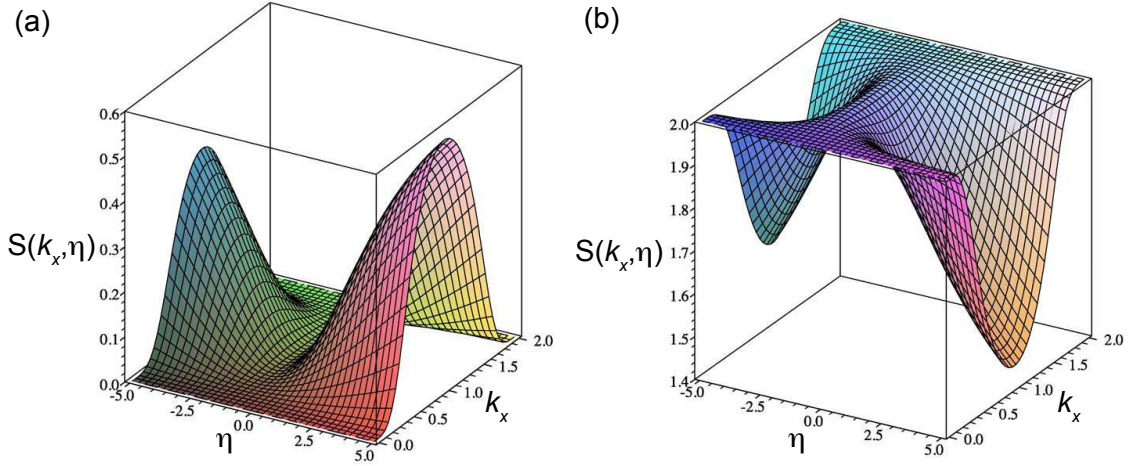}
\caption{$S(k_x,\eta)$ for the (a) high- and (b) low-frequency modes of the FM state with $\gamma$ = 3 and $k_y$ = 0.}
\label{3Dplot-kx}
\end{figure}

\subsection{The FM Phase}

Both the SW frequencies and SF intensities of the FM phase can be obtained analytically. 
The SW frequencies are described by
\be
\begin{array}{c}
\omega_{\mathbf{k}} = {\it B} +JS\Big(2\,{\it \gamma}+\left( \eta -1
 \right)( \cos \left( {\it k_x} \right) - 1)
+ R_{1\mathbf{k}}^{\pm}\Big),
\end{array}
\ee
where 
\be
R_{1\mathbf{k}}^{\pm} = \pm \sqrt{(\eta+1)^2(1-\cos(k_x))^2+4\gamma^2\cos(k_y)^2}
\ee
and the lattice constant $a$ has been set to 1. In the limit $k_x$ = 0,
$\omega_{\mathbf{k}} = {\it B} +2\gamma JS\big(1 \pm \cos \left( {\it k_y}
 \right) \big)$ only depends on $\gamma J$, which is consistent with previous
 results \cite{sas:92,fis:04a,fis:04b,fis:04c}.
 As a function of $\gamma$, $\eta$, and $\mathbf{k}$, the SF intensities are given by
\be
W_{\mathbf{k}} = \frac{R_{1\mathbf{k}}^{\pm} +2\gamma \cos(k_y)}{R_{1\mathbf{k}}^{\pm}}.
\ee
This describes the SF intensities for both the low (+) and high (-) frequency
SW. 

Figure \ref{3Dplot-kx} shows $S(k_x,\eta)$ with $\gamma$ = 3 and $k_y$ = 0 for the high (a)
and low (b) SW modes of the FM phase. If $\gamma$ is held constant, the plot forms a saddle, where the intensity at $\eta$ = -1 is the maximum
throughout $k_x/\pi$. If $\gamma~ \ge$ 1, then this maximum is constant over $k_x$, which is consistent with the standard two-dimensional FM. When $\gamma~<$ 1, the maximum displays non-linear behavior as $k_x$ goes to $\pi$.
The sharpness of the saddle depends
on $\gamma$. As $\gamma$ approaches 0, the saddle sharpens to a delta function at $\eta$ = -1, while increasing $\gamma$ flattens the saddle. The mode with the maximum intensity is also determined by $\gamma$. If $\gamma~>$ 0, then the low-frequency mode dominates the intensity. The opposite is true for negative $\gamma$. As $\eta$ departs from -1 the SF intensity becomes more dependent on $k_x/\pi$ and the intensity decreases.
Therefore, even though the system is a two-dimensional FM,
the non-equivalent interactions create a unique intensity pattern. When the intensities are summed over the high and low frequencies, the total intensity is constant throughout $\mathbf{k}$. This demonstrates that the complex interactions distribute the intensity between the two SW branches, but the total intensity still remains constant as in the standard ferromagnet.

\subsection{The SC1 Phase}

The SC1
region is described by angles ($\theta_a$,$\theta_b$) that are given by
Eq. (\ref{Eminperp}). The SW frequency coefficients for the GVM are given by Eq. (\ref{AntiCoef})
 in Appendix B. In this
section, we present the SW frequencies and SF intensities for varying $\gamma$,
$\eta$, and $B^{\prime}$ to give a representative cross-section of the CAF region.
Due to the complexity of the SF intensities, they have been determined numerically. 

\begin{figure}
\includegraphics[width=5.5in]{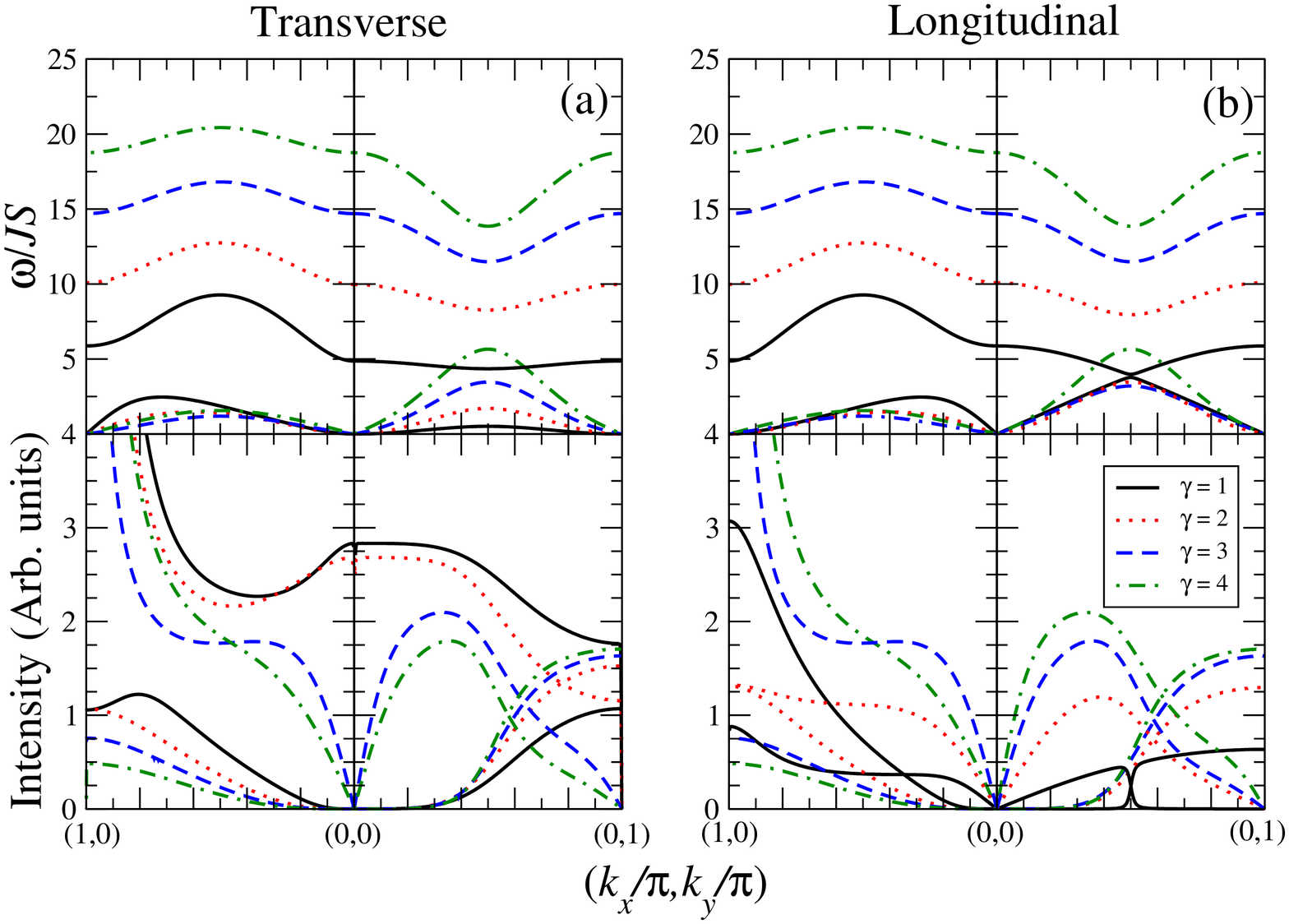}
\caption{(a) Transverse low- and high-frequencies SW modes at $B^{\prime}$ = 0 and $\eta$ = 4
for $\gamma$ = 1...4 as function of ($k_x/\pi$,$k_y/\pi$). Their corresponding SF intensities over the
same values are shown below. The higher-frequency modes have a lower intensity in the $k_x$ direction. (b) Longitudinal modes and intensities activated by the canting of the local moment.}
\label{SW-SF-xy-04gamma}
\end{figure}

The region $\gamma$ = 1 has been examined in great detail in 
Refs. [\onlinecite{sas:92}] and [\onlinecite{fis:04b}], 
where the SW intensities
increase dramatically as ($k_x/\pi$,$k_y/\pi$) approaches (1,0). A more moderate shift
in the intensity is seen along (0,$k_y/\pi$). As shown below, similar features 
are observed as $\gamma$ is increased.  

Fig. \ref{SW-SF-xy-04gamma} shows the low and high SW frequencies and SF 
intensities for the longitudinal and transverse modes at zero field with $\eta$ = 4 and
$\gamma$ = 1...4. The lower SW frequencies tend to decrease in the ($k_x/\pi$,0) direction and 
increase in the (0,$k_y/\pi$) direction. This difference aries because the exchange $\gamma J$ is along the $y$ direction. 
As $\gamma$ increases, the shift in the
intensities along the (0,$k_y/\pi$) direction become more pronounced. As $\gamma$ decreases, 
the SW intensity along the (0,$k_y/\pi$) direction disappears as the FM and AF chains become decoupled.

As shown in Fig. \ref{VillainPhase}(a), the corners
of the zero-field phase diagram consist of two planar
regions sketched in Fig. \ref{VillainPhase} (b) and (c).
A magnetic field causes the spins in both regions to cant. In Fig. \ref{SW-SF-xy-04gamma},
we investigate the transition from the SC1 phase into this planar regime.
While the SW frequencies do
not show much overall difference, the SF intensities do
show a distinct change that would help distinguish between the canted SC1 phase and its
planar limit. The change seen in Fig. \ref{SW-SF-xy-04gamma} clearly shows
 the loss of intensity for the SW modes at (0,0) after entering 
the planar region with $\gamma~>$  8/3. Note that the transverse and longitudinal frequencies become identical
in the planar phase. But for $\gamma~<$ 8/3 in the canted phase the SW modes are either purely transverse or longitudinal. 

\begin{figure}
\includegraphics[width=5.5in]{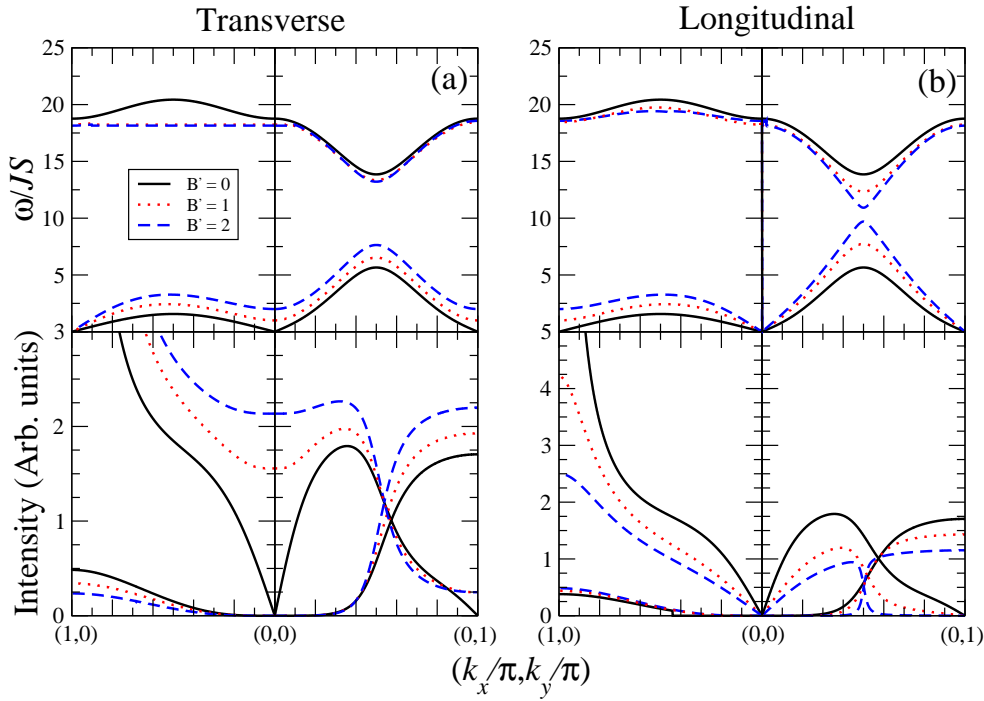}
\caption{(a) Transverse low- and high-frequency SW modes for $B^{\prime}$ = 0...2
with $\eta$ = $\gamma$ = 4 as function of ($k_x/\pi$,$k_y/\pi$).  The corresponding SF intensities over
the same values are shown below. The higher frequency modes typically have a lower intensity in the $k_x$ direction. (b) Longitudinal modes and intensities activated by the canting of the local moments.}
\label{SW-SF-xy-B44}
\end{figure}

Once a magnetic field is applied, the SW frequencies increase due to the enhanced
stiffness of the local moment. Figure \ref{SW-SF-xy-B44} shows the progression
of the high and low SW frequencies and their corresponding SF intensities as the applied magnetic field
 increases from $B^{\prime}$ = 0...2  with $\eta$ = $\gamma$  = 4. This transition
is clearly evident in the low-frequency intensity at (0,0), which becomes non-zero as the field increases and the
spins cant towards the $z$-axis.

\subsection{The SC2 Phase}

The angle $\alpha$
was given terms of the field and $\gamma$ by Eq. (\ref{alpha}). Although $\alpha$ does not depend on $\eta$, 
changing $\eta$ does modify the SW frequencies and SF intensities. The SW frequencies for
this phase can be determined analytically by the equation given by Fishman \cite{fis:04b}
assuming different coefficients (Eq. (\ref{parallelCoef}) and (\ref{SWF}) in Appendix B). As with the
SC1, the SF intensities must be determined
numerically. However, as for the FM phase, the zero-field intensities of the SC2 phase with $\alpha$ = $\pi$/2
can be solved analytically.

In zero field, the SW frequencies are
\be
\begin{array}{c}
\omega_{\mathbf{k}} = \sqrt {R_{2\mathbf{k}}^+R_{2\mathbf{k}}^-} \pm
\left( \eta+1 \right)(1-\cos \left( {\it k_x} \right)),
\end{array}
\label{SC2freq}
\ee
where
\be
R_{2\mathbf{k}}^{\pm}=(\eta-1 ) (\cos( {\it k_x} ) -1)+2\,{\it \gamma}(-1 \pm \,\cos ( {\it k_y} )).
\ee
The SF intensities are given analytically by
\be
W_{\mathbf{k}} =\sqrt{\frac{R_{2\mathbf{k}}^+}{R_{2\mathbf{k}}^-}}.
\ee
Even though the angles of the local moments do not change throughout
the zero-field phase, the SW frequencies and SF intensities depend on the 
interaction parameters, $\gamma$ and $\eta$, where the zero-field intensity for each 
branch in Eq. (\ref{SC2freq}) is the same.

\begin{figure}
\includegraphics[width=5.5in]{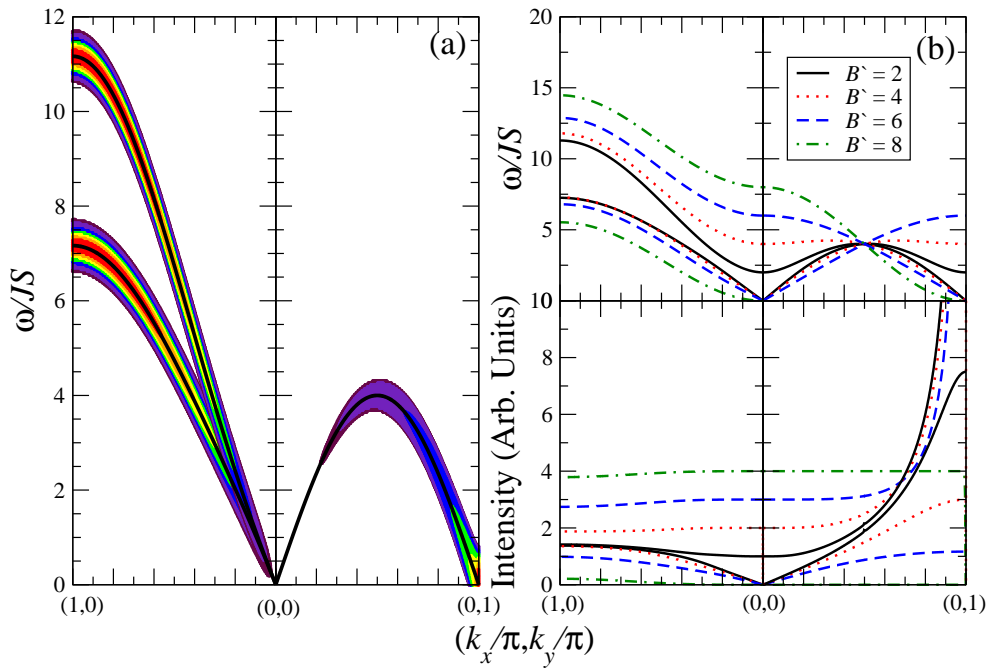}
\caption{(a) Contour plot of the low- and high-frequency SW modes for $B^{\prime}$ = 0
with $\eta$ = $\gamma$ = -2 as function of ($k_x/\pi$,$k_y/\pi$). SF intensity is indicated from to
highest to lowest (red - indigo). Note that the intensity of the $k_x$ is on a linear scale,
while $k_y$ intensity is on a logarithmic scale due to a strong peak at (0,1).(b) Low- and high-frequency SW modes for $B^{\prime}$ = 0...8 with $\eta$ = $\gamma$ = -2 as function of ($k_x/\pi$,$k_y/\pi$). Corresponding SF intensities over the same values. The higher-frequency modes have a lower intensity in the $k_x$ direction.}
\label{Cont-xy-0-2-2}
\end{figure}

Figure \ref{Cont-xy-0-2-2}(a) shows a contour plot of the SW frequencies and SF
intensities for the zero-field SC2 phase with $\gamma$ = $\eta$ = -2. The diagram in the right-hand corner,
illustrates the zero-field angles for the SC2 phase, where the local moments have angles
 $\alpha$ = $\pi$/2. The intensity of the 
modes in the ($k_x/\pi$,0) region are on a linear scale, while the
(0,$k_y/\pi$) intensity is set on a logarithmic scale with an intensity peak at (0,1). 

To further investigate the field-induced canting in this phase, we evaluate the SW frequencies
and SF intensities for various
magnetic field for $\gamma$ = $\eta$ = -2 (shown in Fig. \ref{Cont-xy-0-2-2}(b)). This demonstrates
the transition from the planar SC2 phase into a field-induced canted phase and then finally
into the FM phase as $\alpha$ goes from $\pi$/2 to 0. Each mode of the SC2 phase has 
both longitudinal and transverse character. This is different from the SC1 phase, where the 
longitudinal modes are separate from the transverse.

\section{Conclusion}

In an attempt to understand the nature of competing FM and AF interactions and canted
spin moments, we
present a new technique for the determination of the static and dynamic properties for any
periodic magnetic spin system. Using Euler angles, we
determine the interactions within the local frame of reference for each spin, and
apply these rotations to a Holstein-Primakoff expansion to determine the classical energy and
 SW frequencies. A Green's function technique is used to determine the SF intensities for 
 any eigenfrequency of the system. 

This technique was then applied to the
generalized Villain model  (GVM), which has been further generalized by introducing a varying interchain coupling. By studying the affects of this new interaction, we hope  to gain deeper understanding of frustrated magnetic
systems. To obtain the phase
boundaries and dynamics throughout the GVM$^{\prime}$, we introduce two spin configurations:
SC1 and SC2.

With the phase space established, the SW frequencies and SF intensities for the
different phases were determined. The SW frequencies were determined 
analytically through the whole phase space.
In most cases, the SF intensities were determined numerically. However, we were
able to provide an analytical equation for the FM and zero-field SC2 phase that showed how the SF intensity
depends on the interactions. Using these quantities, we presented a cross-section
of the phase space to give an overall picture of the GVM$^{\prime}$.

We hope that this technique will prove useful 
in the understanding of many different non-collinear magnetic systems. Magnetic 
heterostructures as well as frustrated AFs like CuFeO$_2$ are
some of the systems that may be studied with this technique.

\section{Acknowledgements}

We would like to acknowledge helpful conversations with
Satoshi Okamoto and Wayne Saslow. This research was sponsored by the Laboratory
Directed Research and Development Program of Oak
Ridge National Laboratory, managed by UT-Battelle, LLC
for the U.S. Department of Energy under Contract No. DEAC05-
00OR22725, and by the Division of Materials Science.

\appendix

\section{Rotation Coefficients}

Using Euler rotations with angles ($\theta$ and $\psi$) \cite{arf:01}, each spin $\mathbf{S}_i$ is rotated into its local frame of reference using $\mathbf{\bar{S}}_i$ = $\underline{U}_i\mathbf{S}_i$, where the rotation matrix, $\underline{U}_i$, is given by
\be 
\underline{U}_i = 
 \left| \begin{array}{ccc}
\cos(\theta_i)\cos(\psi_i) & \cos(\theta_i)\sin(\psi_i)  & -\sin(\theta_i) \\
-\sin(\psi_i) & \cos(\psi_i)& 0 \\
\sin(\theta_i)\cos(\psi_i) & \sin(\theta_i)\sin(\psi_i) & \cos(\theta_i) \end{array} \right|.
\label{rotationmatrix}
\ee
When examining the interaction between two spin operators, the
overall rotation from one reference frame to another is
\be
\underline{U}_i\underline{U}_j^{-1} =  \left| \begin{array}{ccc}
F_{xx}^{ij} & F_{xy}^{ij} & F_{xz}^{ij} \\
F_{yx}^{ij} & F_{yy}^{ij} & F_{yz}^{ij} \\
F_{zx}^{ij} & F_{zy}^{ij} & F_{zz}^{ij} \\ \end{array} \right|,
\ee
where $F_{\alpha\beta}^{ij}$ can be obtained from Eq. (\ref{rotationmatrix}). It should be noted that there are $s_t$ different unitary matrices, one for each magnetic SL $r$.
The rotation coefficients for the generalized second-order Hamiltonian
(Eq. (\ref{GenHam})) are
\be
\begin{array}{c}
G_1^{rs} = -\frac{1}{2}(F_{xx}^{rs} + F_{yy}^{rs} - i(F_{xy}^{rs} - F_{yx}^{rs})) \\ \\
G_2^{rs} = -\frac{1}{2}(F_{xx}^{rs} - F_{yy}^{rs} - i(F_{xy}^{rs} + F_{yx}^{rs})).
\end{array}
\ee

When calculating the SW intensities, one needs to multiply each element in the spin Green's function matrix by a rotation coefficient that describes the transverse and longitudinal contributions to the SF. The rotational coefficients of the transverse components
$\big< \mathbf{S}_r^{\pm} (0)\mathbf{S}_s^{\mp} (t)\big>$ are given
by
\be
\begin{array}{c}
\displaystyle 2C_{\pm,\mp}^{2s-1,2r-1} = C_{xx,xx}^{rs} +C_{yx,yx}^{rs} - (C_{xy,xy}^{rs}+C_{yy,yy}^{rs}) \mp (C_{xx,yy}^{rs}-C_{yx,xy}^{rs}) \mp (C_{xy,xy}^{rs}-C_{yy,xx}^{rs})\\
\displaystyle +i \big(\mp(C_{xx,yx}^{rs}-C_{yx,xx}^{rs})\pm(C_{xy,yy}^{rs}-C_{yy,xy}^{rs})-(C_{xx,xy}^{rs}+C_{yx,yy}^{rs})-(C_{xy,xx}^{rs}+C_{yy,yx}^{rs}) \big) \\ \\
\displaystyle 2C_{\pm,\mp}^{2s,2r-1} = C_{xx,xx}^{rs} +C_{yx,yx}^{rs} - (C_{xy,xy}^{rs}+C_{yy,yy}^{rs}) \pm (C_{xx,yy}^{rs}-C_{yx,xy}^{rs}) \pm (C_{xy,xy}^{rs}-C_{yy,xx}^{rs})\\
\displaystyle +i \big(\mp(C_{xx,yx}^{rs}-C_{yx,xx}^{rs})\pm(C_{xy,yy}^{rs}-C_{yy,xy}^{rs})+(C_{xx,xy}^{rs}+C_{yx,yy}^{rs})+(C_{xy,xx}^{rs}+C_{yy,yx}^{rs}) \big) \\ \\
\displaystyle 2C_{\pm,\mp}^{2s-1,2r} = C_{xx,xx}^{rs} +C_{yx,yx}^{rs} + (C_{xy,xy}^{rs}+C_{yy,yy}^{rs}) \mp (C_{xx,yy}^{rs}-C_{yx,xy}^{rs}) \pm (C_{xy,xy}^{rs}-C_{yy,xx}^{rs})\\
\displaystyle +i \big(\mp(C_{xx,yx}^{rs}-C_{yx,xx}^{rs})\mp(C_{xy,yy}^{rs}-C_{yy,xy}^{rs})-(C_{xx,xy}^{rs}+C_{yx,yy}^{rs})+(C_{xy,xx}^{rs}+C_{yy,yx}^{rs}) \big) \\ \\
\displaystyle 2C_{\pm,\mp}^{2s,2r} = C_{xx,xx}^{rs} +C_{yx,yx}^{rs} + (C_{xy,xy}^{rs}+C_{yy,yy}^{rs}) \pm (C_{xx,yy}^{rs}-C_{yx,xy}^{rs}) \mp (C_{xy,xy}^{rs}-C_{yy,xx}^{rs})\\
\displaystyle +i \big(\mp(C_{xx,yx}^{rs}-C_{yx,xx}^{rs})\mp(C_{xy,yy}^{rs}-C_{yy,xy}^{rs})+(C_{xx,xy}^{rs}+C_{yx,yy}^{rs})-(C_{xy,xx}^{rs}+C_{yy,yx}^{rs}) \big),
\end{array}
\ee
where $C_{ab,cd}^{rs}$ $\equiv$ $U_{ab}^{-1r}U_{cd}^{-1s}$ is defined in terms of the inverse of the rotation matrix elements. Note that $r$ and
$s$ denote a specific SL dependent on angles $\theta$ and $\psi$. The rotational coefficients for the 
longitudinal component $\big< \mathbf{S}_r^{z} (0)\mathbf{S}_s^{z} (t)\big>$ are given by
\be
\begin{array}{c}
\displaystyle 2C_{z,z}^{2s-1,2r-1} = C_{13,13}^{rs} - C_{23,23}^{rs} - i(C_{13,23}^{rs}+C_{23,12}^{rs}) \\ \\
\displaystyle 2C_{z,z}^{2s,2r-1} = C_{13,13}^{rs} + C_{23,23}^{rs} + i(C_{13,23}^{rs}+C_{23,12}^{rs})\\ \\
\displaystyle 2C_{z,z}^{2s-1,2r} = C_{13,13}^{rs} + C_{23,23}^{rs} - i(C_{13,23}^{rs}+C_{23,12}^{rs}) \\ \\
\displaystyle 2C_{z,z}^{2s,2r} = C_{13,13}^{rs} - C_{23,23}^{rs} + i(C_{13,23}^{rs}+C_{23,12}^{rs}).
\end{array}
\ee

\section{Spin-Wave Frequency Coefficients}

When $\gamma$ = 1, the second order Hamiltonian (Eq. (\ref{GVMHam})) and SW frequencies were 
determined analytically by Fishman in Ref. [\onlinecite{fis:04b}].
For the GVM$^{\prime}$ Hamiltonian with general $\gamma$, the coefficients for SC1 are
\be
\begin{array}{c}
A_{\mathbf{k}}^{(a,a)} = 2\cos(2\theta_a) + 2\gamma\cos(\theta_a - \theta_b) -2\cos^2(\theta_a)\cos(k_x) + B^{\prime}\cos(\theta_a) \\ \\
A_{\mathbf{k}}^{(a,b)} = A_{\mathbf{k}}^{(b,a)} =  - 2\gamma\cos^2((\theta_a-\theta_b)/2)\cos(k_y) \\ \\
A_{\mathbf{k}}^{(b,b)} = 2\gamma\cos(\theta_a-\theta_b) - 2\eta\cos(2\theta_b) + 2\eta\cos^2(\theta_b)\cos(k_x) + B^{\prime}\cos(\theta_b) \\ \\
B_{\mathbf{k}}^{(a,a)} = \sin^2(\theta_a)\cos(k_x) \\ \\
B_{\mathbf{k}}^{(a,b)} = B_{\mathbf{k}}^{(b,a)} = - 2\gamma\sin^2((\theta_a-\theta_b)/2)\cos(k_y) \\ \\
B_{\mathbf{k}}^{(b,b)} = -\eta\cos^2(\theta_b)\cos(k_x),
\end{array}
\label{AntiCoef}
\ee
while the coefficients for SC2 are given by
\be
\begin{array}{c}
A_{\mathbf{k}}^{(a,a)} = 2(1 + \gamma\cos(\theta_a - \theta_b) - \cos(k_x)) + B^{\prime}\cos(\theta_a) \\ \\
A_{\mathbf{k}}^{(a,b)} =A_{\mathbf{k}}^{(b,a)} =  - 2\gamma\cos^2((\theta_a-\theta_b)/2)\cos(k_y) \\ \\
A_{\mathbf{k}}^{(b,b)} = 2(\gamma\cos(\theta_a-\theta_b) - \eta + \eta\cos(k_x)) + B^{\prime}\cos(\theta_b) \\ \\
B_{\mathbf{k}}^{(a,a)} = 0 \\ \\
B_{\mathbf{k}}^{(a,b)} = B_{\mathbf{k}}^{(b,a)} = - 2\gamma\sin^2((\theta_a-\theta_b)/2)\cos(k_y) \\ \\
B_{\mathbf{k}}^{(b,b)} = 0.
\end{array}
\label{parallelCoef}
\ee

The SW frequencies are given in terms of $A_{\mathbf{k}}^{(r,s)}$ and $B_{\mathbf{k}}^{(r,s)}$ by
\be
\omega_{\mathbf{k}} = \frac{JS}{\sqrt{2}}\sqrt{A_{\mathbf{k}}^{(a,a)2} + A_{\mathbf{k}}^{(b,b)2}
+2(A_{\mathbf{k}}^{(a,b)2}-B_{\mathbf{k}}^{(a,b)2}) - 4(B_{\mathbf{k}}^{(a,a)2}+B_{\mathbf{k}}^{(b,b)2}) \pm R_{3\mathbf{k}}}
\label{SWF}
\ee
\be
\begin{array}{c}
R_{3\mathbf{k}}^2 = 4\Big(A_{\mathbf{k}}^{(a,a)2}+A_{\mathbf{k}}^{(b,b)2}-4\big(B_{\mathbf{k}}^{(a,a)2}+B_{\mathbf{k}}^{(b,b)2}\big)\Big)\Big(A_{\mathbf{k}}^{(a,b)2}-B_{\mathbf{k}}^{(a,b)2}\Big) + \Big(A_{\mathbf{k}}^{(a,a)2}-A_{\mathbf{k}}^{(b,b)2}-4\big(B_{\mathbf{k}}^{(a,a)2}-B_{\mathbf{k}}^{(b,b)2}\big)\Big)^2 \\ \\
+8\Big(A_{\mathbf{k}}^{(a,a)}A_{\mathbf{k}}^{(b,b)}+4B_{\mathbf{k}}^{(a,a)}B_{\mathbf{k}}^{(b,b)}\Big)\Big(A_{\mathbf{k}}^{(a,b)2}+B_{\mathbf{k}}^{(a,b)2}\Big) - 32A_{\mathbf{k}}^{(a,b)}B_{\mathbf{k}}^{(a,b)}\Big(A_{\mathbf{k}}^{(a,a)}B_{\mathbf{k}}^{(b,b)}+A_{\mathbf{k}}^{(b,b)}B_{\mathbf{k}}^{(a,a)}\Big)
\end{array}
\ee


\begin{thebibliography}{harald}

\bibitem{kit:87} Kittel C
{\it Quantum Theory of Solids}
(Wiley, 1987).

\bibitem{vil:77} Villain J 1977 {\it J. Phys. C} {\bf 10} 1717.

\bibitem{ber:86} Berge B, Diep H T, Ghazali A, and Lallemand P
1986 {\it Phys. Rev. B}
{\bf 34} 3177.

\bibitem{khr:00} Khrapai V S, Deviatov E V, Shashkin A A, Dolgopolov V T,
Hastreiter F, Wixforth A, Campman K L,
and Gossard A C 2000 {\it Phys. Rev. Lett.} {\bf84} 725.

\bibitem{san:01} S‡nchez D, Brey L, and
Platero G. 2001 {\it Phys. Rev. B} {\bf 64} 235304.

\bibitem{laz:04} Lazarovits B, Ujfalussy B, Szunyogh L, Stocks GM, and
Weinberger P 2004 {\it J. Phys. Cond. Mat.} {\bf 16} S5833.

\bibitem{dui:03} van Duijn J, Attfield J P, Watanuki  R, Suzuki K, Heenan R K 2003
{\it Phys. Rev. Lett.} {\bf90} 087201.

\bibitem{cof:91} Coffey D., Rice T. M, and Zhang F C 1991
{\it Phys. Rev. B} {\bf 44} 10112.

\bibitem{she:93} Shekhtman L, Aharony A, and Entin-Wohlman O
1993 {\it Phys. Rev. B} {\bf 47} 174.

\bibitem{jor:01} Jorgensen J D, Chmaissem O, Shaked H, Short S, Klamut P W,
Dabrowski B, and Tallon J L 2001 {\it Phys. Rev. B} {\bf 63} 054440.

\bibitem{cao:01} Cao G, McCall S, Zhou Z X, Alexander C S,
Crow J E, Guertin R P, and Mielke C H 2001 {\it Phys. Rev. B} {\bf 63} 144427.

\bibitem{nak:02} Nakamura K. and Freeman A. J. 2002 {\it Phys. Rev. B} {\bf 66} 140405(R).

\bibitem{dag:01} For an overview of phase separation in the manganites,
see Moreo A, Yunoki S, and Dagotto E 1999 {\it Science} {\bf 283} 2034
and Dagotto E, Hotta T, and Moreo A 2001 {\it Phys. Rep.} {\bf 233} 1.

\bibitem{ada:00} Adams C P, Lynn J W, Mukovskii Y M, Arsenov A A, and Shulyatev D
2000 {\it Phys. Rev. Lett.} {\bf 85}, 3954.

\bibitem{koo:01} Koo T Y, Kiryukhin B, Sharma P A, Hill J P, and Cheong S-W 2001
{\it Phys. Rev. B} {\bf 64} 220405.


\bibitem{wal:80} Walker L R and Walstedt 1980 {\it Phys. Rev. B} {\bf 22} 3816.

\bibitem{sas:83} Saslow W M 1983 {\it Phys. Rev. B} {\bf 27} 6873.

\bibitem{sas:92} Saslow W M and Erwin R 1992 {\it Phys. Rev. B} {\bf 45} 4759.

\bibitem{car:04} Carlson E W, Yao D X, and Campbell D K 2004 {\it Phys. Rev. B} {\bf 70} 064505.

\bibitem{fis:04a} Fishman R S 2004 {\it Phys. Rev. B} {\bf70} 012403.

\bibitem{fis:04b} Fishman R S 2004 {\it J. Phys. Cond. Mat.} {\bf 16} 5483.

\bibitem{fis:04c} Fishman R S 2004 {\it Phys. Rev. B} {\bf70} 140402(R).



\bibitem{arf:01} Arfken G B and Weber H J,
{\it Mathematical Methods for Physicists}
(Academic Press, 2001).

\bibitem{squ:78} Squires G L
{\it Introduction to the Theory of Thermal Neutron Scattering}
(Dover, 1996).

\bibitem{bal:89} Balcar E and Lovesey S W
{\it Theory of Magnetic Neutron and Photon Scattering}
(Oxford, 1989).

\bibitem{gab:89} Gabay M, Garel T, Parker G N, and Saslow W M 1989 {\it Phys. Rev. B}
{\bf 40} 264.

\end{thebibliography}
\end{document}